\documentclass[a4paper,oneside,12pt]{report}
\usepackage{styles/fbe_tez}
\usepackage[utf8x]{inputenc} 
\usepackage{amsmath, amsthm} 
\usepackage[bottom]{footmisc}
\usepackage{cite}
\usepackage{graphicx}
\usepackage{longtable}
\usepackage{float}
\usepackage{multirow}
\usepackage{subfigure}
\usepackage{algorithm}
\usepackage{algorithmic}
\usepackage{array}
\usepackage{amssymb,bm,cite,graphicx, fixmath, texdraw}
\usepackage{epsfig}
\usepackage{epstopdf}
\usepackage{subfigure} 
\usepackage{amsmath}
\usepackage{notoccite}
\usepackage{mathrsfs}
\usepackage{xcolor}

 \makeatletter

\newcommand{\beq}{\begin{equation} \setlength\abovedisplayskip{5pt} 
\setlength\belowdisplayskip{5pt}}
\newcommand{\eeq}{\end{equation}}
\newcommand{\bea}{\begin{eqnarray}}
\newcommand{\eea}{\end{eqnarray}}

\def\ifundefined{\@ifundefined}
\def\bfat{\left[ \begin{array}}
\def\emat{\end{array} \right]}
\def\bfatt{\left\{ \begin{array}}
\def\ematt{\end{array} \right.}
\def\bset{\left\{ \begin{array}}
\def\eset{\end{array} \right\}}
\def\bpar{\left( \begin{array}}
\def\epar{\end{array} \right)}

\graphicspath{{figures/}} % Graphics will be here

\numberwithin{equation}{chapter}

\title{ON THE DEVELOPMENT OF THE CONCEPT OF THE WEAK COSMIC CENSORSHIP CONJECTURE IN GENERAL RELATIVITY}
\turkcebaslik{GENEL GÖRELİLİKTEKİ ZAYIF KOZMİK SANSÜR VARSAYIMI KAVRAMININ GELİŞİMİ ÜZERİNE}
\degree{B.S., Physics, Yıldız Technical University, 2015}
\author{Ahmet Cem Erdoğan}
\program{Physics}
\subyear{2023}

\supervisor{Prof. İbrahim Semiz}
\examineri{Assoc. Prof. Koray Düztaş}
\examinerii{Assoc. Prof. Hakan Erkol}
\dateofapproval{17.11.2023}

\begin{document}
\pagenumbering{roman}
\makemstitle % M.S. thesis
\makeapprovalpage

% ACK PAGE
\begin{acknowledgements}
First, I would like to thank my thesis advisor, İbrahim Semiz, for his support, guidance, and unwavering attention. His insights into a wide range of subjects have been an encouragement to me, and I have learned a multitude of valuable lessons from him. 

I would like to express my appreciation to Koray Düztaş for guiding me. His advice and suggestions on the subject were very helpful. Thanks to him, I was able to move forward very quickly on the points I was stuck on.

Another thank you to Ercüment Akat who asked \emph{Cevizlibağ} questions. His help is undeniable. With his always open library and the books he gave me to read on the \emph{Marmaray}, he both increased my motivation and made me think deeper about physics.

 I cannot thank my wife and son enough. The support they give me in terms of time is unbelievable. They have shown me love, respect, and understanding. Even though my son would crawl under the chair during my work and wait for me to accompany him with various musical instruments, I am deeply grateful to him.

I would also like to thank my grandmother, Bedia. Her help and support, though seemingly small to her, meant a lot to me. Without her, I believe this work would not have been possible. Finally, I want to express my gratitude to my entire family for their support.
\end{acknowledgements}

% ABS PAGE
\begin{abstract}
%In this thesis, we trace the evolution of the definition of the weak cosmic censorship conjecture and the attempts to prove or disprove it. This thesis involves delving into the development of this conjecture, examining its conceptual transformations, and scrutinizing the various attempts made to validate or invalidate the weak cosmic censorship conjecture. This scholarly exploration seeks to shed light on the complexities surrounding the weak cosmic censorship conjecture, ultimately contributing to a deeper understanding of this fundamental concept within the realm of classical general relativity. In the first chapter, the development of thoughts about singularities resulting from gravitational collapse is briefly introduced. In the second chapter, causality and related mathematical concepts are discussed. In the third chapter, the reasons for proposing weak cosmic censorship are introduced. In the fourth and fifth chapters, literature reviews from the 1970s and 1980s are provided.

In this thesis, we scrutinize the literature to trace the evolution of the concepts surrounding the weak cosmic censorship conjecture and the attempts to prove or disprove it. We discuss its development over the years,  its conceptual transformations, and the various thought experiments constructed to validate or invalidate the weak cosmic censorship conjecture. We also discuss possible connections with blackhole thermodynamics, and end with an evaluation of the current status about the subject, and finally, speculations about the future.
\end{abstract}

% OZET PAGE
\begin{ozet}
%Bu tezde, zayıf kozmik sansür varsayımının tanımının evriminin ve onu kanıtlama ya da çürütme girişimlerinin izini sürüyoruz. Bu tez, bu varsayımın gelişimini araştırmayı, kavramsal dönüşümlerini incelemeyi zayıf kozmik sansür varsayımını doğrulamak veya geçersiz kılmak için yapılan çeşitli girişimleri irdelemeyi içermektedir. Bu bilimsel araştırma, zayıf kozmik sansür varsayımını çevreleyen karmaşıklıklara ışık tutmayı ve nihayetinde klasik genel görelilik alanındaki bu temel kavramın daha derinlemesine anlaşılmasına katkıda bulunmayı amaçlamaktadır. Birinci bölümde, kütleçekimsel çökme sonucu oluşan tekillikler hakkındaki düşüncelerin gelişimi kısaca tanıtılmaktadır. İkinci bölümde nedensellik ve ilgili matematiksel kavramlar tartışılmaktadır. Üçüncü bölümde, zayıf kozmik sansürü önermenin nedenleri tanıtılmaktadır. Dördüncü ve beşinci bölümlerde 1970'ler ve 1980'lerdeki literatür taramalarına yer verilmiştir.
Bu tezde, zayıf kozmik sansür varsayımını çevreleyen kavramların evrimini ve onu kanıtlama veya çürütme girişimlerini izlemek için literatürü inceliyoruz. Yıllar içindeki gelişimini, kavramsal dönüşümlerini ve zayıf kozmik sansür varsayımını doğrulamak veya geçersiz kılmak için inşa edilen çeşitli düşünce deneylerini tartışıyoruz. Ayrıca karadelik termodinamiği ile olası bağlantıları tartışıyor, konuyla ilgili mevcut durumun bir değerlendirmesi ve son olarak gelecekle ilgili spekülasyonlarla bitiriyoruz.

\end{ozet}

% CONTENTS AND LIST OF FIG AND TABLE PAGES
\tableofcontents
\listoffigures
%\listoftables

% LIST OF SYMBOLS PAGE
\begin{symbols}
% The title will be typeset as "LIST OF SYMBOLS".
% Use a separate \sym command for each symbols definition.
% First, Latin symbols in alphabetical order
%\sym{$a_{ij}$}{Description of $a_{ij}$}

\sym{$a$}{Total angular momentum of a black hole per unit mass}
\sym{$A_h$}{The surface area of the event horizon of a black hole}
\sym{c}{The speed of light in vacuum}
\sym{$D^+(S)$}{Future domain of dependence}
\sym{$D^-(S)$}{Past domain of dependence}
\sym{$D(S)$}{Full domain of dependence}
\sym{G}{Gravitational constant}
\sym{$g_{\mu\nu}$}{Metric tensor with the signature $(-,+,+,+)$}
\sym{$g'_{\mu\nu}$}{Metric tensor in a new coordinate system}
\sym{$H^+(S)$}{Future Cauchy horizon}
\sym{$H^-(S)$}{Past Cauchy horizon}
\sym{$H(S)$}{Full Cauchy horizon}
\sym{$I^+(p)$}{Chronological future of event p}
\sym{$I^-(p)$}{Chronological past of event p}
\sym{$J$}{Total angular momentum of a black hole}
\sym{$J^+(p)$}{Causal future of event p}
\sym{$J^-(p)$}{Causal past of event p}
\sym{$M$}{Mass of a black hole}
\sym{$\mathbf{M}$}{Manifold}
\sym{$p$}{Pressure of a perfect fluid}
\sym{$R$}{Ricci scalar}
\sym{$R_{\mu\nu}$}{Ricci tensor}
\sym{$R_{\mu\nu\rho\sigma}$}{Riemann tensor}
\sym{$q$}{Electric charge of a black hole}
\sym{$T_p$}{Tangent space at point $p$}
\sym{$T_{\mu\nu}$ }{Energy-momentum tensor}
\sym{$V_h$}{Electric potential of a black hole at the event horizon}
\\
\sym{$\epsilon_0$}{Electric permittivity of free space}
\sym{$\kappa$}{The surface gravity of a black hole}
\sym{$\Lambda$}{Cosmological constant}
\sym{$\rho$}{Energy density for a perfect fluid}
\sym{$\rho_m$}{Mass density for a perfect fluid}
\sym{$\Sigma$}{Achronal surface}
\sym{$\Omega^2$}{Conformal factor}
\sym{$\Omega_h$}{Angular velocity of a black hole at the event horizon}

\\
\sym{$\hbar$}{Reduced Planck constant}
\sym{ $\mathcal{I}^+$}{Future timelike infinity}
\sym{ $\mathcal{I}^-$}{Past timelike infinity}
\sym{ $\mathcal{I}^0$}{Spacelike infinity}
\sym{$\mathscr{I}^+$}{Future null infinity}
\sym{$\mathscr{I}^-$}{Past null infinity}
\sym{$\mathcal{L}$}{Lagrangian density}

\end{symbols}

% LIST OF ABBREV PAGE
\begin{abbreviations}
 % Abbreviations in alphabetical order
%\sym{2D}{Two Dimensional}
%\sym{3D}{Three Dimensional}
%\sym{AAM}{Active Appearance Model}
\sym{DEC}{Dominant Energy Condition}
\sym{EFE}{Einstein Field Equation}
\sym{EHC}{Event Horizon Conjecture}
\sym{GEC}{Generic Energy Condition}
\sym{GR}{General Relativity}
\sym{NDEC}{Null Dominant Energy Condition}
\sym{NEC}{Null Energy Condition}
\sym{OSD}{Oppenheimer-Snyder-Datt}
\sym{SEC}{Strong Energy Condition}
\sym{WCCC}{Weak Cosmic Censorship Conjecture}
\sym{WEC}{Weak Energy Condition}
\end{abbreviations}

\chapter{INTRODUCTION}
\label{chapter:introduction}
\pagenumbering{arabic}
Until the 1960s, black holes were considered solely as mathematical solutions to the Einstein Field Equations due to their inclusion of singularities. The occurrence of a singularity was considered very strange, leading Einstein to believe that black holes could not exist as real astronomical objects. In the 1960s, Stephen Hawking, Roger Penrose, and Robert Geroch introduced singularity theorems that included causality conditions, which became crucial to preserving the causal structure of spacetime. Accepting singularities within the context of general relativity means acknowledging the limitations of the theory while simultaneously preserving the causal structure of spacetime.

\section{Singularity}
In the realm of contemporary scientific understanding, we possess a sophisticated theory to elucidate the workings of the universe in a classical framework. Einstein formulated his groundbreaking, renowned \textquotedblleft General Theory of Relativity" in 1915. Initially, he thought that the field equation associated with this theory could not be exactly solved\cite{einstein1916}. Nevertheless, one year later, Karl Schwarzschild exactly solved the field equation for the case of a static, spherically symmetric, uncharged, non-rotating body. Interestingly, Einstein had already provided and used an approximate solution\footnote{The differential line element is  $ds^2=-(1-\frac{2GM}{c^2r})c^2dt^2+(1+\frac{2GM}{c^2r})dr^2+r^2(d\theta^2+\sin^2\theta d\phi^2)$} to explain the peculiar motion of Mercury's perihelion\cite{earman1995}. While analyzing the solution, he discovered the existence of two singularities but he consciously abstained from delving into their meaning because they were not the main problem. However, after acquiring Schwarzschild's exact solution, the presence of singularities became significantly problematic due to the fact that scientist sought to predict the future events based on the initial events. 
\newpage
Consequently, Hilbert proposed a definition in order to explain the meaning of these singularities. According to Hilbert's definition\footnote{This definition implies that the concept of a metric is about metric “potentials" in a coordinate system. However, in GR, a metric is an intrinsic object; it exists independently of coordinate systems.\cite{earman1999}\cite{eisenstaedt1993}}\!\!\!, the metric tensor $g_{\mu\nu}$ is not  singular at a point if it is possible to introduce  $g_{\mu\nu}^{'}$ which is a new metric tensor in a new coordinate system. The coordinate transformation from $g_{\mu\nu}$ to $g_{\mu\nu}^{'}$ must be reversible, one-to-one transformation such that in the new coordinate system the corresponding metric $g_{\mu\nu}^{'}$ is not singular at that point and it must be continuous and differentiable at the point and in a neighborhood of the point. Additionally, the determinant $g'$ is required to not be equal to zero.

%{\footnote{In today's literature, it is recognized that Hilbert's definition of singularity contains an error. There exists a coordinate system that is regular; however, under a coordinate transformation, it can become singular. For example, let's consider the cartesian coordinate system. In this coordinate system, no singular point exists. Yet, upon changing the coordinate system from cartesian to cylindrical coordinates, a singular point arises at $r=0$. Consequently, the definition is fundamentally flawed.}}

Based on his definition, one can easily conclude that the Schwarzschild metric is singular at both the center of the spherical body ($r=0$)  and the Schwarzschild radius ($r=r_s=\frac{2GM}{c^2}$ where $G$ is universal constant, $\mathbf{M}$ is the mass of the spherical body and $c$ is the speed of light in vacuum). At these points, the metric is not continuous and differentiable, therefore they are singular. Einstein thought that Schwarzschild's metric does not represent a physical metric and constructed some arguments to explain his thought. 

 In 1932, Georges Lemaitre proposed an explanation for the singularities in Schwarz-schild metric. His idea was an explicit and self-consistent demonstration. He calculated a curvature invariant called “Kretschmann curvature invariant” $K=R_{\mu\nu\rho\sigma}R^{\mu\nu\rho\sigma}$ for the singular points. At the center of the sphere, this  scalar blows up, whereas at the Schwarzschild radius, the Kretschmann curvature invariant remains well behaved. Thus, the singularity at the Schwarzschild radius must be a coordinate singularity in the light of Lemaitre's proposition.  There was no reason to believe a singularity at $r=r_S$ as a genuine singularity. By employing Hilbert's definition, the search for a coordinate system that does not give a genuine singularity in the Schwarzschild radius was initiated. In 1960, Kruskal and Szekeres published a paper on the maximal extension for the Schwarzshild solution. This effectively put an end to the debate regarding the singularity at the Schwarzschild radius.

However, despite considerable progress,  several questions regarding singularities remained unanswered. What does the singularity at r=0 signify? Is it a genuine or a coordinate singularity? What exactly is a spacetime singularity? Until 1960s, the exact answer to these questions was not clear. 

The concept of singularity exists within the realm of classical physics as well. The distinction between classical and relativistic physics in understanding the concept of singularity is significant. To illustrate this distinction, we can examine the notion of singularity in electromagnetic theory. In this context, if the electromagnetic energy density reaches an infinite value at a point, a singularity is said to exist at that point. However, within the framework of general relativity, we are not only concerned with singularities in the field but also with singularities in spacetime itself.

Hence, it becomes necessary to establish a precise definition of what a singularity is. Mathematically, it is a difficult concept to define. If we define a singular point as a point where the curvature (Riemann tensor) of spacetime blows up, we acknowledge that the metric behaves inconsistently at that specific point. Consequently, we can confidently state that the point is not a part of spacetime since the metric of spacetime is ill-defined at that point. As an alternative approach, similar to Lemaitre, we examine the curvature scalars and observe that for singular points, the curvature scalar becomes infinitely large. Nevertheless, this approach also presents its own set of problematic situations.\footnote{For plane gravitational waves, all curvature scalars can vanish  even though the Riemann tensor is not regular.}

For the third attempt, we consider geodesic incompleteness, as singularities involve ill behavior in curvature, which should be related to geodesically incomplete curves. A geodesic curve is said to be incomplete if it possesses finite affine length. Consequently, a spacetime is said to be singular if it is geodesically incomplete. There are spacetime observers that may reach the end of spacetime if spacetime is geodesically incomplete. They can disappear from the universe in a finite time. 

Roger Penrose, Stephen Hawking and Robert Geroch tried to explain the singularity in a spacetime. Here is the theorem given by Penrose and Hawking\cite{hawkingpenrose1970} to define the singularity.

Let $\mathbf{M}$, $g_{\mu \nu}$ be a time-oriented spacetime satisfying the following four conditions:
 \begin{enumerate}
 \item $R_{\mu \nu} V^{\mu}V^{\nu} \geq 0$ for any non-spacelike $V^{\mu}$.
 \item The timelike and null generic conditions are fulfilled.
 \item There is no closed timelike curve.
 \item At least one of the following holds:
 \begin{enumerate}
 \item There exists a compact achronal set without edge.
 \item There exists a trapped surface.
 \item There  is a $p \in \mathbf{M}$ such that the expansion of the future (or past) directed null geodesic through $p$ becomes negative along each of the geodesics.
 \end{enumerate}
 \end{enumerate}
Then $\mathbf{M}$, $g_{\mu \nu}$ contains at least one incomplete timelike or null geodesic.

This theorem does not use physical laws; it has a purely geometrical meaning. Thus, some energy conditions are required to explain the meaning of singularity. These energy conditions will be introduced in the second chapter of the thesis.
\section{Cosmic Censorship}
After the development of the singularity theorems, Penrose considered the possibility that under some circumstances the event horizon of a black hole might disappear and black hole without event horizon might be found in the universe. If so, all physical quantities will blow up and a timelike-moving particle or object can suddenly appear out of the singularity and enter the spacetime. Thus, one can lose the power of the prediction and this causes a very problematic situation. Penrose called the singularity inside the black hole that has no event horizon as naked. Naked singularity is the singularity which can be observed by an external observer.

The idea regarding the cosmic censorship conjecture was first introduced by Roger Penrose in 1969. Penrose asked the question: “Does there exist a `cosmic censor' who forbids the appearance of naked singularities, clothing each one in an absolute event horizon?”. During that time, the existence of big bang singularity was recognized and the hot Big Bang was verified by cosmic microwave backround. However, the big bang singularity is an exception for a naked singularity. Penrose asserted that we cannot know the cause of the big bang singularity or whether singularities arise from perfectly reasonable non-singular initial data. He proposed a hypothesis that every black hole must hide their naked singularities clothed by their event horizons so that the outside spacetime is protected from the possibly causality-violating effects of the singularity. This hypothesis is called the weak cosmic censorship hypothesis. He did not state a mathematical definition for this hypothesis.

Penrose, Hawking, Wald, Israel and many other notable physicists emphasized the conjecture as the most important unresolved challenge within classical general relativity. To validate this conjecture, several definitions have been proposed, and various approaches have been undertaken. Although many brilliant physicists showed intense effort to prove the hypothesis, the weak cosmic censorship conjecture has remained as a hypothesis. Since 1969, a lot of research papers have been published regarding the problematic conjecture. Interestingly, the change of formulation of the weak cosmic censorship conjecture took form around attemped counter-examples.
\section{Outline of the thesis}

In the literature, several examples and counter-examples included unrealistic idealizations, physically impossible models can be found regarding the validation of weak cosmic censorship hypothesis. The reason behind these examples and counter-examples is the evolving nature of the definition associated with the conjecture. It is likely to confuse new researchers who are about to start research on this conjecture. The objective of this thesis is to trace the progression of the definition of weak cosmic censorship and attempt to provide guidance for further research in this area. In this thesis, the ideas associated with the evolution of the concept of the weak cosmic censorship conjecture will be studied.

\chapter{CAUSALITY IN GR AND SOME DEFINITIONS}
\label{chapter:causality in gr and some definition}
In this chapter, we will introduce some definitions utilized to understand causality in general relativity. Additionally, causality condition and violation will be reviewed.
\section{Time-Orientability}
\label{Time-Orientability}
Before the definition of the time orientability, we must define vectors in relativistic physics. By defining the metric $g_{\mu\nu}$, we have the inner product for an arbitrary four vector ${\widetilde{V}}$ with itself as
\beq
{\widetilde{V}} \cdot {\widetilde{V}}=g_{\mu\nu}V^{\mu}V^{\nu}=V^{\mu}V_{\mu}=V^{\nu}V_{\nu}.
\eeq 
A vector $\widetilde{V}$ is said to be 
\begin{enumerate}
\item a time-like vector if ${\widetilde{V}} \cdot {\widetilde{V}}<0$,
\item a space-like vector if ${\widetilde{V}} \cdot {\widetilde{V}}>0$,
\item a light-like vector if ${\widetilde{V}} \cdot {\widetilde{V}}=0$,
\item  a causal vector if ${\widetilde{V}} \cdot {\widetilde{V}}\leq 0$.
\end{enumerate}
 
 Consider a spacetime with $\mathbf{M}$ and $g_{\mu\nu}$ and select an event $p\in M$. We have a tangent space $T_p$ at the point $p$, in which there exist time-like vectors divided into two different classes. The designation of the classes is a choice of a direction for the arrow of time due to the increase of entropy. Hence, these two classes are labeled future-directed and past-directed. At the point $p$, we have both future-directed and past-directed vectors.  In case where the designation of time-like vectors as future or past directed vectors can be continuously made over the entire $\mathbf{M}$ globally, this space-time is said to be time-orientable. This means, in general, the time-like direction field can be replaced by a time-like vector field if the space-time is time-orientable. 
 
Time-orientability is vitally important not only mathematically, but also physically. Our models used to elucidate the universe must be time-orientable since there exists an agreement on the future time direction among observers in different regions of spacetime. 
%For a different perspective on time-orientability, one can examine  elementary particle physics within the frame of symmetry violations (ya açıkla ya da çıkar).
% A lorentzian manifold is time-orientable manifold.
\section{Causal Curves}
\label{Causal Curves}
Let $\gamma(\lambda)$ be a curve with the parameter $\lambda$. A tangent vector is given by
\beq
\gamma\,'(\lambda)=\frac{d\gamma}{d\lambda}.
\eeq
If the tangent vector of curve $\gamma\,'(\lambda)$ is a time-like vector everywhere, the curve $\gamma$ is called a time-like curve. It can also be referred to as a worldline. If the tangent vector of curve $\gamma\,'(\lambda)$ is everywhere a light-like vector, the curve $\gamma$ is called a light-like or null curve. If the tangent vector of curve $\gamma\,'(\lambda)$ is a space-like vector everywhere, the curve $\gamma$ is called a space-like curve. A curve is causal (non space-like) if the tangent vector of the curve $\gamma\,'$ is time-like or light-like. 

%Conditions on the tangent vectors of the curves define the causal relationships in a spacetime.

\section{Pasts and Futures}
\label{Pasts and Futures}
 Let a point $p$ be an event on a space-time. The light signal emerging at event $p$ forms a light cone.\footnote{These light cones are the fundamental invariant structure in Minkowski spacetime. There does not exist an analogy of a light cone in Euclidean space.} Since the upper speed limit is the speed of light,  causally related events may occur within the light cone originating from event $p$. If this space-time is time-orientable, two light cones exist in that space-time, where the one of the cones represents the future, the other represents the past. The light cone at point $p$ is a subset of $T_p \, \mathbf{M}$. Then, there are two disjoint open sets of time-like vectors with one future pointing  and one past pointing lightcone. 
 \begin{figure}[htbp]
 \centering
 \includegraphics[scale=.6]{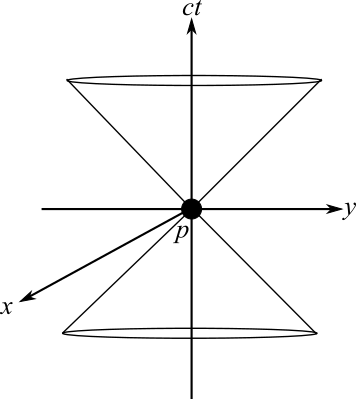}
  \caption{Light cone.}
		\label{fig:light cone}
 \end{figure}

Consider that $\widetilde{V}$  is a timelike vector and $\widetilde{W}$ is a non-zero causal vector, $\widetilde{W}$ is future pointing, $W^+$, if \textcolor{white}{jjjjjkjbh}
\beq
 g_{\mu\nu}\, V^{\mu}W^{\nu}<0.
\eeq Similarly, $\widetilde{W}$ is past pointing, $W^-$, if \textcolor{white}{jjjjjkjbh}
\beq
 g_{\mu\nu}\, V^{\mu}W^{\nu}>0.
\eeq

A timelike curve $\gamma(\lambda)$ is said to be future/past directed if the tangent vector of the curve is  future/past pointing at each point $p\in\gamma$. If the tangent vector of the curve is a future/past directed timelike or null vector, $\gamma(\lambda)$ is said to be future/past directed causal curve. A curve $\gamma(\lambda)$ is said to be future/past inextendible if the curve has no future/past end point.

 The chronological future/past of an event $p$ in spacetime is defined as the set of events which can be reached by a future/past directed timelike curve originating from $p$. Denoted as $I^+(p)$ for the chronological future and $I^-(p)$ for the chronological past, these sets hold significant physical implications. $I^-(p)$ represents the collection of all events of spacetime which can influence what happens at $p$ while $I^+(p)$ encompasses the collection of all events of spacetime which can be influenced by what happens at $p$. Both $I^+(p)$ and $I^-(p)$ are open sets in spacetime, excluding their boundaries. Notably, the boundaries of $I^+(p)$ and $I^-(p)$ correspond to the future and past null cones, respectively.
 \newpage
  Let $p$ and $q$ be two events in spacetime ($\mathbf{M}$ and $g_{\mu\nu}$) where $p\in \mathbf{M}$ and $q\in \mathbf{M}$. Event $q$ can influence event $p$ if and only if $q$ is in $I^-(p)$. This means that $q$ can be joined to $p$ with future-directed timelike curve if and only if $p$ can be joined to $q$ with past-directed timelike curve. Intersection of $I^-(p)$ and $I^+(q)$, $I^-(p)\cap I^+(q) $, is called causal diamond. No closed timelike curves exist in a causal diamond.
 
  The causal future/past of an event $p$ in spacetime is defined as the set of events which can be reached by a future/past directed causal curve starting from $p$. The causal future of $p$ is denoted by $J^+(p)$ whereas the causal past of $p$ is denoted by $J^-(p)$. Both $J^+(p)$ and $J^-(p)$ are closed sets in Minkowski spacetime but they may not necessarily be closed in general spacetime.
  
%We can generelize these defition to the set of events.
\section{Domain of dependence and Cauchy Horizons}
\label{Domain of dependence and Cauchy Horizons}
%In spacetime, there may exist not only an event but also two or more events. If we study the description of the fundamental relation of influence among events mathematically, we must introduce the subject of the domain of influence.  
  
  Let $S$ be a set of events with $S\subset \mathbf{M}$. The set $S$ is said to be semi-spacelike (achronal) if there are no two points in $S$ connected by a timelike curve, which means that there do not exist two points $p$ and $q\in S$ such that $p \in I^+(q)$. Namely, the set $S$ is achronal if $I^+(S)\cap S=\emptyset$. To illustrate this, we can consider an edgeless spacelike hypersurface in Minkowski spacetime. This hypersurface is achronal, which means that it might be a spacelike or null three-dimensional surface.
  
  \begin{figure}
  \centering
  \includegraphics[scale=.5]{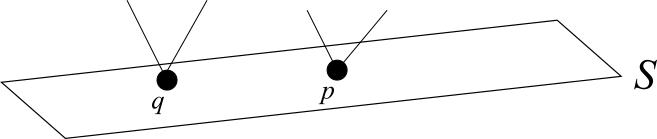}
  \caption{Achronal surface.}
  \label{fig:Achronal Surface}
  \end{figure}
  
%   The chronological future of the set of events, $I^+(S)$, is called the future domain of influence of $S$. We can denote the past domain of influence of $S$ as $I^-(S)$, analogously.
 
The question has been asked: Is the behavior at event $p$ entirely determined by what occurs on a spacelike, three-dimensional surface $S$? To answer this question we must introduce the notion of domain of dependence. In fact, the concept of the domain of dependence stands in dual relationship to that of the domain of influence.
\newpage
 Consider a spacelike three-dimensional surface $S$ in a time-orientable manifold $\mathbf{M}$ with the condition $S\in \mathbf{M}$. Let there be an event $p$ in the future of the surface $S$. Information propagates along null or timelike curves, affecting specific regions influenced by event $p$ rather than all regions. Thus, we define the future or past domain of dependence of $S$ as $D^+(S)$ or $D^-(S)$, respectively. This definition helps us determine the region where past or future-directed timelike and null curves from event $p$ intersect the surface $S$. The full domain of dependence is the union of $D^+(S)$ and $D^-(S)$, which is \textcolor{white}{jjjjjkjbhgcvhgchgc}
\beq
D(S)=D^+(S)\cup D^-(S).
\eeq
Since $D^+(S)$ and $D^-(S)$ are closed, $D(S)$ must be closed. The boundary of $D^+(S)$ is called as the future Cauchy horizon, denoted by $H^+(S)$. The past Cauchy horizon, $H^-(S)$ , can be defined by exchanging the future and the past. The future/past Cauchy horizon is defined by \textcolor{white}{jjjjjkjbhgcvhgchgc}
\beq
H^{\pm}(S)=\partial D^{\pm}(S).
\eeq
 As the future and past Cauchy horizon are the boundaries of $D^+(S)$ and $D^-(S)$, they are null surfaces. This implies that the future and past Cauchy horizons are achronal, meaning that there are no points on these horizons that give rise to past or future directed null curves within them. The full Cauchy horizon is defined by
 \beq
 H(S)=H^+(S)\cup H^-(S).
 \eeq
\section{Cauchy Surface}
\label{Cauchy Surface}
A Cauchy surface is defined by the achronal surface $\Sigma$ if the full domain of dependence of the achronal surface $\Sigma$ equals the entire manifold $\mathbf{M}$ as
\beq
D(\Sigma)=\mathbf{M}.
\eeq
 Since a closed, edgeless achronal set is a three-dimensional  embedded $C^0$ submanifold of $\mathbf{M}$,  every Cauchy surface is an embedded $C^0$ submanifold of $\mathbf{M}$. In other words, a spacetime with a Cauchy surface must be of the form $S^3 \times \mathbb{R}$. Thus, we could conceptualize the Cauchy surface as symbolizing an instantaneous snapshot of time throughout the universe. If we know the Cauchy surface of spacetime, we can predict all possible events in the universe. Spacetime is said to be globally hyperbolic if spacetime possesses a Cauchy surface. 

Not all exact solutions of EFE possess a Cauchy surface. Schwarzschild and Friedmann spacetimes possess a Cauchy surface while maximally extended Reissner-Nordström and Kerr spacetimes do not\footnote{In our universe, Reissner-Nordström and Kerr spacetimes possess a Cauchy surface.}\cite{hawkingisrael1979}.
\section{Causality Violation and Conditions}
\label{Causality Violation and Conditions}
The general theory of relativity states that spacetime must be locally Minkowski spacetime. Since Minkowski spacetime is time-orientable, it exhibits both future and past light cones. These cones specify which events may be causally related to each other through timelike or light signals, defining the causal structure in the spacetime. Consequently, an event in spacetime is always preceded by its cause. This defines the causality condition locally.

However, defining the causality condition is more complicated when dealing with global topologies. In curved spacetime, the causality condition will be different from that of Minkowski spacetime. Light cones behave differently, and this can cause confusion if we use the causality condition from Minkowski spacetime. There is a need for a causality condition in global topologies.

A closed curve is a curve whose endpoints are the same point. A closed timelike curve is a closed curve that is everywhere future-directed timelike (or everywhere past-directed timelike). A closed null curve is a closed curve that is everywhere future-directed null (or everywhere past-directed null). If we consider a spacetime that includes closed causal curves, this implies that we need to define additional conditions beyond the laws of physics to accurately describe reality. Suppose that we conduct some experiments in spacetime. Since closed timelike curves exist, the initial conditions of the experiment can be influenced again during the experiment. The implication is that closed causal curves cause the causality violation.

Another implication of causality violation is the compactness of spacetime. Every compact spacetime leads to a causality violation\cite{hawkingisrael1979}. Therefore, we need to define the conditions under which spacetime excludes both closed causal curves and compact spacetime. These conditions are referred to as causality conditions.

A spacetime is strongly causal if $\forall$ $p\in \mathbf{M}$ and for every neighborhood $O$ of $p$, there exists a neighborhood $N\subset O$ containing $p$ such that no causal curve $\gamma$ intersect the neighborhood $N(p)$ more than once. If the spacetime is strongly causal, the spacetime also is globally hyberbolic. However, strong causality is not a good condition for the causality due to the fact that it admits almost closed timelike or null curves. 

A spacetime is said to be stably causal if and only if there exists a differentiable function $f$ on $\mathbf{M}$ such that $\nabla^af$ is a past directed timelike vector field. The concept of stable causality carries the implication of strong causality, but stably causal spacetimes rule out the possibility of the existence of closed timelike or null curves. From the definition, if a spacetime is stably causal, it implies that there is a time function, namely, a slice that is admitted by every stably causal spacetime. These slices might be achronal, but achronal slices are not necessary for stably causal spacetimes. There exist some spacetimes that are stably causal but do not include achronal slices.

 If a spacetime possesses a Cauchy surface, it must exhibit stable causality\footnote{In contrast, a spacetime is not required to possess a Cauchy surface if it is stably causal.} \!\!\!. Therefore, there exists a time function such that achronal slices at constant time are also Cauchy surfaces of the spacetime. All of these slices must also be diffeomorphic, i.e. topologically identical. Schwarzschild, Reissner-Nordström, Friedmann and Weyl spacetimes satisfy the condition of stable causality.
 \begin{figure}
\centering
\includegraphics[scale=.4]{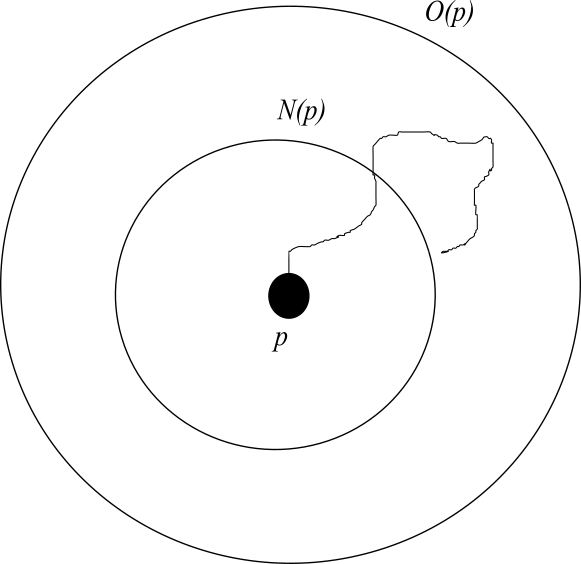}
\caption{Strong causality.}
		\label{fig:strong causality}
\end{figure}
\section{Energy Conditions}
\label{Energy Conditions}
Einstein Field Equations (EFE) with the cosmological constant $\Lambda$ is given by
\beq
R_{\mu\nu}-\frac{1}{2}g_{\mu\nu}\,R+ \Lambda\, g_{\mu\nu}=\frac{8\pi\, G}{c^4}T_{\mu\nu}
\eeq
The left-hand side of the equation pertains to the geometry whereas the right-hand side corresponds to the energy-momentum tensor i.e. matter and energy. From that equation, we consider all the metrics obey EFE unless we have some restrictions on the energy-momentum tensor $T_{\mu\nu}$. In the physical context, we must impose restrictions on the energy-momentum tensor to model realistic sources of energy and momentum, as there is no inherent restriction to differentiate between the mathematical and physical solutions. The following conditions are not related to energy conservation as Bianchi identity ensures that the covariant derivative of energy-momentum tensor equals zero. 

%Thus, additional restrictions of $T_{\mu\nu}$ are not related to the energy conservation.

 It is necessary to impose energy conditions that are coordinate-invariant restrictions. Thus, we construct scalars using $T_{\mu\nu}$. Some suggested energy conditions are defined as follows: Weak Energy Condition (WEC) states that 
\beq
T_{\mu\nu}V^{\mu}V^{\nu}\geq 0
\eeq 
for all timelike vectors $V^{\mu}$. This implies that the local energy density must be positive when it is measured by any timelike observer. Null Energy Condition (NEC) states that \textcolor{white}{easilyjbkjbnkjbnkjbkjbıhnhkıj}
\beq
T_{\mu\nu}L^{\mu}L^{\nu}\geq 0
\eeq for all null vectors $L^{\mu}$. It is a generalization of the WEC including lightlike vectors. Strong Energy Condition (SEC) states that
\beq
T_{\mu\nu}V^{\mu}V^{\nu} \geq \frac{1}{2}V^{\mu}V_{\mu} T
\eeq
 for all timelike vectors $V^{\mu}$ ($\Lambda=0$). Dominant Energy Condition (DEC) includes WEC and it has an additional requirement, which is 
\beq
T_{\mu\nu}T_{\sigma}^{\nu}V^{\mu} V^{\sigma}\leq0
\eeq
for all timelike vectors $V^{\mu}$. Namely, $T^{\mu\nu}V_{\mu}$ is a nonspacelike (causal) vector. Physically, this condition implies that the local energy density must be positive (due to the WEC) and the energy flux must be timelike or lightlike. Both electromagnetic and scalar fields obey this condition. Furthermore, the WCCC is considered assuming the DEC.

The Null Dominant Energy Condition (NDEC) is similar to the DEC, but it differs in that it involves null vectors instead of timelike vectors in the DEC. NDEC states that NEC is valid and \textcolor{white}{easilyhjhjjhjh}
\beq
T_{\mu\nu}T_{\sigma}^{\nu}L^{\mu} L^{\sigma}\leq0
\eeq
for all null vectors $L^{\mu}$. Namely, $T^{\mu\nu}L_{\mu}$ is a nonspacelike (causal) vector.

To illustrate these conditions physically, one can consider the energy-momentum tensor for a perfect fluid. The energy-momentum tensor for a perfect fluid is given by
\beq
T^{\mu\nu}= (\rho_m+p/c^2)u^{\mu}u^{\nu}+g^{\mu\nu}p
\eeq
where $u^{\mu}$ is four-velocity vector field, $\rho_m$ is total mass density of the fluid and $p$ is pressure. We may denote the total energy density of the perfect fluid by $\rho=\rho_m c^2$.

 WEC implies that $\rho\geq 0$ and $\rho +p\geq 0$, while SEC satisfies the conditions $\rho+p\geq 0$ and $\rho +3p\geq 0$. Namely, the energy density of the fluid cannot be negative for the WEC. For the SEC, energy density might be negative or there might be large negative pressure. For the NEC, we have $\rho +p\geq 0$, and this means the energy density might be negative. For the DEC, we have  $\rho\geq |p|$. For the NDEC, the negative energy densities are allowed so long as $p=-\rho$.
 
Given the belief in an inflationary epoch after the Big Bang, the cosmological constant may play a vital role in these suggested energy conditions. If $\Lambda \neq 0$, the total energy-momentum tensor is expressed by
\beq
T^{\mu\nu}_{\text{total}}=T^{\mu\nu}-\frac{\Lambda}{8\pi G}g^{\mu\nu}.
\eeq
In the case of $\Lambda > 0$, the WEC, NEC, and DEC are satisfied, but the SEC is violated. In the case of $\Lambda < 0$, the NEC and SEC are satisfied, while the WEC and DEC are violated. Additionally, we have Generic Energy Condition (GEC). It is valid if
 \begin{enumerate}
 \item SEC holds.
 \item Every timelike or null geodesic contains a point where $L_{[\mu}R_{\nu]\rho\sigma[\alpha}L_{\beta]}L^{\rho}L^{\sigma}\neq 0$ 
 \end{enumerate}
 
If the GEC holds, every geodesic path will pass through an area where gravitational focusing occurs. This implies that pairs of conjugate points\footnote{Conjugate points can be joined by a 1-parameter family of geodesics. For example, the north and south poles on a sphere can correspond to conjugate points.} exist if one can extend the geodesic far enough in each direction. However, the GEC is not satisfied by some known exact solutions of the EFE.
 
The validity of the energy conditions is a reasonable assumption for classical physics. Therefore, energy conditions are invoked in the singularity theorems. However, many physical systems (especially quantum systems) are known to violate energy conditions. Pion and Higgs fields, which are minimally coupled massive scalar fields, can violate the strong energy condition. The Lagrangian density of these type of fields is given by \textcolor{white}{jjjjjkjbhgcvhgchgcsadasdsad}
\beq
\mathcal{L}=\frac{1}{2}\,[m^2\phi^2+(\nabla \phi)^2].
\eeq
The energy-momentum tensor of the field becomes
\beq
T^{\mu\nu}=\nabla^{\mu}\phi\nabla^{\nu}\phi-\frac{1}{2}\,g^{\mu\nu}\,[m^2\phi^2+(\nabla \phi)^2].
\eeq 
Thus, one can easily violate the SEC by taking $\phi$ as a time-independent scalar function with the frame $V^{\mu}=(c,0,0,0)$.

Moreover, the Casimir effect, which demonstrates the distortion of the zero-point energy of the vacuum between imposed boundaries, provides experimental evidence that all the energy conditions can be violated. While the WEC and DEC are violated due to possible negativity of energy density\footnote{In fact, Casimir energy is not required to be negative, it can be positive for some boundary geometries\cite{casimir1988}.}\!\!\!, the NEC can also be violated with a quick calculation. Since the NEC can be violated, the SEC can also be violated. These violations are not large; they are very small, on the order of $\hbar$ effects.
\chapter{WEAK COSMIC CENSORSHIP CONJECTURE}
\label{WEAK COSMIC CENSORSHIP CONJECTURE}
The Oppenheimer-Snyder-Datt (OSD) model stands as the most idealized representation within the realm of gravitational collapse utilizing the EFE. The OSD model gives the final result for the gravitational collapse of a  pressureless, homogeneous massive star. Namely, this model represents a formation scenario for a Schwarzschild black hole. Two important issues arise for this model: the singularity at the center of the black hole and the event horizon of the black hole. The singularity at the center of a black hole is the focus of the singularity theorems. The event horizon of the black hole shapes the spacetime uniquely compared to classical physics. If an object or light crosses the event horizon of a black hole, it cannot return. Even light cannot escape after passing an event horizon. To be more precise, the nature (timelike vs spacelike) of the $t$ coordinate (the coordinate with respect to which the spacetime is translationally symmetric) and the radial coordinate are exchanged at the event horizon. The singularity at the center of a black hole is called naked unless the event horizon of the black hole forms.
\section{Naked Singularity}
\label{Naked Singularity}
The implication of naked singularities would be catastrophic if they exist. Since there is no event horizon, particles and light can escape from the strong gravity region. The curvature of spacetime around a singularity is large; therefore, local physics could completely change around naked singularities. Namely, local determinism can break down if naked singularities exist, by particles/waves emerging from singularity. This means that one could not say anything about future evolution of an event. Hence, the concept of event horizon might isolate the gap between the concept of singularity and determinism.
 
Another perspective on naked singularities is as follows: Naked singularities can lead to the expulsion of matter from themselves due to the time-reversal symmetry of the EFE. As there does not exist a preferred time direction in the EFE, there should be a time reversed version of a black hole ( referred to as a white hole). For this case, light cannot reach a white hole but light might emerge from it.

\section{Kerr-Newman Black Hole}
\label{Kerr-Newmann Black Hole}
The unique exact solution of the black holes which are charged and rotating is given by the metric of Kerr-Newmann black hole as
\beq
ds^2 =-\frac{\Delta}{\rho^2}\left[ c\,dt-a\sin^2\theta\, d\phi\right]^2+\frac{\sin^2\theta}{\rho^2}\left[(r^2+a^2)\,d\phi-a\,c\,dt\right]^2+ \frac{\rho^2}{\Delta}\,dr^2+\rho^2 \, d\theta^2,
\eeq 
where $\Delta$ and $\rho^2$ are
\beq
\begin{split}
\Delta=& r^2-\frac{2GMr}{c^2}+\frac{1}{4\pi\epsilon_0}\frac{Gq^2}{c^4}+a^2, \\
\rho^2=& r^2+a^2 \cos^2\theta,
\end{split}
\eeq
where $a=\frac{J}{Mc}$ . Since $\frac{1}{4\pi\epsilon_0}=G=c=1$ in Gaussian units, we have
\beq
\begin{split}
\Delta=& r^2-2Mr+q^2+a^2, \\
\rho^2=& r^2+a^2 \cos^2\theta.
\end{split}
\eeq

The Kerr-Newman black hole, which depicts the gravitational collapse of charged, rotating stars, is one of the exact solutions of EFE. The solution represents asymptotically flat, axisymmetric, stationary \footnote{It cannot be static because it does not have a time reversal symmetry.  It is stationary because of its consistent rotation, which remains unchanged over time.} spacetime for the Einstein-Maxwell equations. At the beginning of the gravitational collapse, the metric of spacetime does not resemble that of a Kerr-Newman black hole. However, after event horizons form, the final state of the gravitational collapse leads to the Kerr-Newman black hole. Two event horizons exist at \textcolor{white}{jjjjjkjbhgcvhgchgc}
\beq \label{eq:3.4}
r_{\pm}=M\pm \sqrt{M^2-a^2-q^2}
\eeq
In the case of $M^2 \leq a^2+q^2$, the real roots of $\Delta$ cannot be found; thus, event horizons do not form.
\newpage
Prior to 1969, Birkhoff's theorem, which implies that the Schwarzschild solution is the unique solution that describes a spherically symmetric, uncharged, non-rotating black hole, was known.  Similar to Birkhoff's theorem, Werner Israel proposed two theorems that claim the stability of the event horizon of all static black holes \cite{israel1967}\cite{israel1968}. Penrose referred to a generalization of these theorems as the Generalized Israel Conjecture (GIC), which states that the gravitational collapse of a rotating, charged mass culminates in a Kerr-Newman black hole.  According to GIC, the event horizon of the Kerr-Newman black hole cannot be unstable and it cannot disappear\footnote{The proof was given by Mazur in 1982 \cite{mazur1982}}. If it does, the singularity at $r=0$ will then be visible for external observers. This implies that in this region there will be no equation capable of predicting the future events since there is no complete Cauchy surface.

Penrose introduced these circumstances in the 1969 paper\cite{penrose1969}. Thus, he proposed the weak cosmic censorship conjecture, which states that every black hole  forming after a gravitational collapse with  reasonable, nonsingular initial data must have an event horizon. The center singularity of the black hole cannot be naked, according to this conjecture.
\section{Conformal Transformation}
\label{Conformal Transformation}
The proposition of weak cosmic censorship is related to the disparity between Penrose's approach in GR emphasizing causal structure\cite{penrose1963} and his contribution to the theory of black holes \cite{penrose1965}. According to Penrose, the concept of an event horizon defines a black hole. However, Penrose did not use the concept of the event horizon in his singularity theorem. Penrose's approach to GR includes conformal transformation of the metric. A conformal transformation is given by
\beq
g'_{\mu\nu}(x)=\Omega^2(x)\, g_{\mu\nu}(x) ,
\eeq
where $\Omega^2$ (called conformal factor) is a positive scalar field. The transformed metric $g'_{\mu\nu}$ may not correspond to a physical metric (It does not necessarily satisfy the EFE.), but the causal structure given by $g'_{\mu\nu}$ is the same as that of $g_{\mu\nu}$\footnote{That means that these two metrics give the same null curves.}. Penrose's idea is that $\Omega$ might be selected to decrease as we approach infinity, in such a way that $g'_{\mu\nu}$ brings infinity within a finite distance. So, by analyzing the behavior of null geodesics in asymptotically flat spacetime, we investigate the global structure of spacetime.  To illustrate this, we may use Minkowski spacetime for the simplicity. The Minkowski metric is given by \textcolor{white}{jjjjjkjbhgcvhgchgc}
 \beq
 ds^2=-c^2dt^2+dr^2+r^2d\Omega^2 ,
 \eeq
 where $d\Omega^2=d\theta^2+\sin^2\theta \, d\phi^2$. The range of coordinates is $-\infty < t < \infty $ and $0 \leq r < \infty $. To extend the range of $r$-coordinates into infinity, we can do substitution $u=ct-r$ and $v=ct+r$ in Minkowski spacetime. By doing this, we actually change the range of coordinates. New coordinate ranges become $-\infty < u < \infty $ and $-\infty < v < \infty $. In fact, the metric becomes
 \beq
 ds^2=-dvdu+\left( \frac{v-u}{2}\right)^2\, d\Omega^2 .
 \eeq
 In order to bring infinity in finite distance, we can do a trasformation as $p=\arctan v$ and $q=\arctan u$. The range of coordinates is $-\frac{\pi}{2}\leq p \leq \frac{\pi}{2}$ and $-\frac{\pi}{2}\leq q \leq \frac{\pi}{2}$. The metric becomes
 \beq
ds^2=-\sec^2 q \sec^2 p\, dpdq-\frac{1}{4}\sec^2 q \sec^2 p \sin^2(p-q)\, d\Omega^2.
 \eeq
 Lastly, substitute $t'=p+q$ and $r'=p-q$. The new ranges will be $-\pi \leq t' \leq \pi$ and $-\pi \leq r' \leq \pi$. The metric will be
 \beq
 ds^2= dt'^2-dr'^2+\sin^2r'\, d\Omega^2.
 \eeq
 
 This metric represents the conformally transformed metric. Implications can be found as follows:
 \begin{enumerate}
 \item As $t\rightarrow \infty$ with finite $r$, $p=q=\pi/2$, $t'=\pi$ and $r'=0$. Thus, this point corresponds to the future timelike infinity denoted by $\mathcal{I}^+$. There is a single endpoint for all future directed timelike geodesics.
  \item As $t\rightarrow -\infty$ with finite $r$, $p=q=-\pi/2$, $t'=-\pi$ and $r'=0$. Thus, this point corresponds to the past timelike infinity denoted by $\mathcal{I}^-$. There is a single endpoint for all past directed timelike geodesics.
  \item As $r\rightarrow \pm \, \infty$ with finite $t$, $t'=0$ and $r'=\pm \, \pi$. Thus, these points correspond to the spacelike infinity denoted by $\mathcal{I}^0$. There is a single endpoint for all spacelike geodesics.
  \item As $t\rightarrow \infty$ and $r\rightarrow \infty$ but $ct-r$ finite, $p=\pi/2$ and $q$ finite, $ 0 \leq t' \leq \pi$ and $ 0 \leq r' \leq \pi$ with $t'+r'=\pi$. This line corresponds to the future null infinity denoted by $\mathscr{I}^+$. All future directed null geodesics end in the future null infinity.
   \item As $t\rightarrow -\infty$ and $r\rightarrow \infty$ but $ct+r$ finite, $q=-\pi/2$ and $p$ finite, $ -\pi \leq t' \leq 0$ and $ 0 \leq r' \leq \pi$ with $t'-r'=-\pi$. This line corresponds to the past null infinity denoted by $\mathscr{I}^-$. All past directed null geodesics end in the past null infinity.
 \end{enumerate}

  In the Penrose diagram for Minkowski spacetime, light rays can be represented by 45-degree lines from the vertical axis; therefore, particles with mass must follow paths that deviate by less than 45 degrees. The chronological future and past of an event $A$ can be seen in the figure. Each point in the Penrose diagram represents a 2-sphere.
   \begin{figure}
  \centering
  \includegraphics[scale=.7]{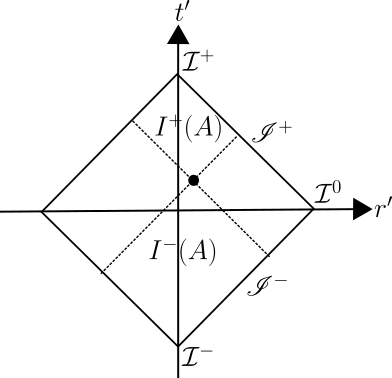}
  \caption{Penrose diagram for an event $A$.}
  \label{fig:Penrose Diagram for an event}
  \end{figure}
  \begin{figure}
  \centering
  \includegraphics[scale=.7]{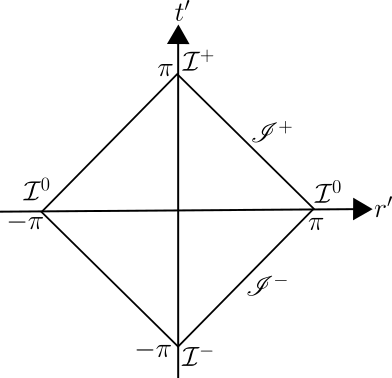}
  \caption{Penrose diagram for Minkowski spacetime.}
  \label{fig:Penrose Diagram for Minkowski spacetime}
  \end{figure}  
  \section{Asymptotic Flatness}
  \label{Asymptotic Flatness}
   A global description of a system can be made in special relativity since there exist Killing vectors in special relativity that lead to the definition of the energy, momentum and angular momentum of the system. However, in general relativity, there are no Killing vectors in the generic\footnote{Generic case means the case without special properties.} case. Thus, defining these quantities in general relativity is challenging. For that reason, the assumption that a spacetime must be asymptotically flat at null infinity can sometimes be used to describe an isolated system in general relativity. In addition to this, we discussed black holes embedded in an asymptotically flat spacetime, although current cosmological models suggest that our universe may not conform to an asymptotically flat configuration. For that reason, we should introduce the definition of an asymptotically flat spacetime. Here is the definition\cite{townsend1997}:

Let $\mathbf{M}$ and $g_{\mu\nu}$ be a time orientable spacetime. This spacetime is asymptotically flat at null infinity if there exists a conformally compactified spacetime $\mathbf{\tilde{M}}$ and $\tilde{g}_{\mu\nu}$ such that
\newpage
\begin{enumerate}
\item Equation(3.5) is satisfied with the condition $d\Omega\neq 0$ on the boundary of the manifold $\partial M$.
\item The manifold $\mathbf{M}$ may be extended to obtain the manifold with its boundary $\mathbf{M}\cup \partial \mathbf{M}$ within the manifold $\mathbf{\tilde{M}}$.
\item The boundary of the manifold $\mathbf{M}$ is the disjoint union of the future and past null infinity and each diffeomorphic to $\mathbb{R} \times S^2$.
\item No future/past directed causal curve starting in the manifold $\mathbf{M}$ intersects the past/future null infinity.
\item $\mathscr{I}^{\pm}$ are complete.
\end{enumerate}

Conditions $(i)$ and $(ii)$ are requirements to get appropriate conformal compactification. Other conditions ensure that the causal structure at infinity are the same as that of Minkowski spacetime.
   \section{Black Hole Thermodynamics}
 \label{Black Hole Thermodynamics}
In 1972, Hawking published a paper\cite{hawking1972} about black holes in which he demonstrated that the size of a black hole's surface area remains constant over time by accepting the stability of the event horizon of a black hole and the weak cosmic censorship conjecture. At the same time, Bekenstein attempted to seek an explanation for the problem posed by Wheeler: What happens when a cup of tea passes through a black hole? Initially, we have a cup of tea with nonzero entropy and a black hole. In the final state, we will have a black hole with a larger area. Thus, the black hole must have an entropy proportional to  the surface area to validate the second law of thermodynamics. The entropy of a black hole \cite{bekenstein1973} is given by
\beq
S \propto \frac{A_h}{G\hbar},
\eeq
where $A_h$ is the area of the event horizon of the black hole. In 1973, Bardeen, Carter, and Hawking \cite{bch1973} proposed four laws of black hole mechanics analogous to the laws of thermodynamics.
\newpage
\begin{enumerate}
\item \emph{Zeroth law of black hole mechanics}: The surface gravity of a stationary black hole (denoted by $\kappa$) does not change over the event horizon of the black hole; it remains constant. The stationary black hole spacetime obeys the DEC.
\item \emph{First law of black hole mechanics}: For a small variation of parameters ${M}$, $J$, $Q$ and $A_h$ \textcolor{white}{jjjjjkjbhgcvhgchgc}
\beq
\delta M= \frac{\kappa}{8\pi G}\delta A_h+ \Omega_h \delta J+V_h \delta Q,
\eeq
where $\Omega_h$ and $V_h$ are the angular velocity and the electric potential at the horizon, respectively.
\item \emph{Second law of black hole mechanics}:The surface area of the event horizon of a black hole (denoted by $A_h$) never decreases such that
\beq
A_h\geq 0.
\eeq
\item \emph{Third law of black hole mechanics}: It is impossible to reduce $\kappa$ to zero by any procedure in a finite number of steps.
\end{enumerate}

The formulation of the second and third laws of black hole mechanics includes the assumption of the validation of the WCCC. They assume that there are no naked singularities in nature.

Since they are analogous to the laws of thermodynamics, they are referred to as the laws of black hole thermodynamics. For the first law of thermodynamics, the energy $E$ and mass $M$ are not analogous; they are the same physical quantity. However, $\kappa$ and $A_h$ cannot correspond to real temperature and real entropy in this context. In classical thermodynamics, heat flows from a cold body to a hot body; however, in the case of a classical black hole, the energy of a heat reservoir will flow into it but never flows out. Nonetheless, an emission of particles is possible in the process of Hawking radiation. Thus, $\kappa$ and $A_h$ can represent physical temperature and entropy when quantum effects are taken into account.
 %\section{Viewpoints on WCC}
 %\label{Viewpoints on WCC}
%According to the WCC, singularities produced by a gravitational collapse from generic, physically reasonable initial data must be hidden. "Generic Einsteinian spacetimes with physically reasonable sources do not admit any naked singularity". "Big bang is not a counter-example to this conjecture it has no past and hence does nıt correspond to any future inextendible causal curve". "Generic is usually taken to meanwithout exceptional properties or stable under small perturbations."
 %stability, causality, gedanken experiments...
 \chapter{EARLY DEVELOPMENT OF THE WCCC IN CLASSICAL GENERAL RELATIVITY}
 \label{Early Development of the WCCC in Classical General Relativity}
 In this chapter, we trace the milestones of the development of the weak cosmic censorship conjecture in classical general relativity from the beginning of the 1970s to the end of the 1980s. 
\section{The Developments of the WCCC in the 1970s}
\label{The Developments of the WCCC in the 1970s}
Investigations of the stability of the event horizon of a black hole were an approach to prove or disprove the WCCC. If the event horizon of a black hole is unstable, the horizon could disappear, thus creating a naked singularity. To prove the conjecture, one can investigate the instabilities of the black hole.

After Penrose's work, first evidence for the weak cosmic censorship conjecture came from Price (1972)\cite{price1972}. Price's approach is to show the stability of the Schwarzschild black hole. His work indicates that gravitational collapse with small departures from spherical symmetry will result in a Schwarzschild black hole.

In 1973, Boulware studied a gravitational collapse scenario for a thin charged shell\cite{boulware1973}. According to this study, a naked singularity forms if and only if the proper mass of the shell is negative. Thus, he emphasized that to prove the weak cosmic censorship conjecture, one must assume positive definite matter density.

In 1973, Yodzis, Seifert, Müller zum Hagen (YSM) provided the first counterexample\cite{ysm1973} to the conjecture.  They showed that a naked singularity might exist for the spherically symmetric dust collapse (OSD works make special choices facilitating mathematical operations during the derivation of their solutions) and for an almost perfect fluid\footnote{YSM called it an almost perfect fluid since the pressure for this model is close to, but not exactly equal to, the perfect fluid pressure.}. However, they did not investigate the stability under perturbation of the equation of state for the almost perfect fluid model, nor did they examine the stability of the dust collapse model. After this work, Penrose said that the weak cosmic censorship conjecture might be false due to several reasons, including this counterexample\cite{penrose1973}. One year later, YSM showed that shell-crossing singularities may exist for spherically symmetric perfect fluids within a large family of equations of state and proved the stability of the solutions\cite{ysm1974}.

In 1974, Robert Wald approached the conjecture from a different angle staring one of the main paths of research about WCCC. He tested the validity of the weak cosmic censorship conjecture by constructing a thought experiment attempting to violate the GIC starting with extremal Kerr-Newman black hole\cite{wald1974}. He took an extremal Kerr-Newman black hole with $M^2=q^2+a^2$ and sent a test particle to create a case with $M^2<q^2+a^2$. For this purpose, the test particle should have a large angular momentum and/or charge compared to its mass. Alternatively, we can send a spinning test particle with a high spin-to-mass ratio. Wald found that the black hole cannot capture test particles in each case. Heuristically, particles with larger angular momentum have too large an impact parameter to be captured by the black hole, particles with large charge are repelled by the black hole; and in the general/combined case, the equation of motion for the test particle guarantees that these effect are combined in just the right manner to prevent the black hole for being pushed beyond extremality.

This work was later generalized by many authors to various kinds (different dimensions, alternative gravity theories, with cosmological constant etc.) of black holes and various types of incoming matter/energy; hence, it can be said to have inspired a significant fraction of the literature on WCCC.

Liang studied the collapse scenario for an irrotational dust collapse\cite{liang1974}. He showed that in the cylindrical case, the collapse leads to a naked singularity. Actually, Thorne independently demonstrated the cylindrical dust collapse leading to a naked singularity\cite{thorne1972}. However, this model is not realistic due to the elimination of the possibility of gravitational waves and cylindrical symmetry\footnote{The extension in the z-direction is infinite.}\!\!\!.

In 1977, Jang and Wald published a paper to eliminate counterexamples of the weak cosmic censorship conjecture\cite{jangwald1977}. Penrose introduced an inequality (referred to as Penrose's inequality) to rule out some possible violations of the weak cosmic censorship conjecture, such as a gravitational collapse of a shell null fluid with flat spacetime inside the shell, in his 1973 paper\cite{penrose1973}. Modifying the positive energy conjecture of Geroch\cite{geroch1973}, they showed that other possible counterexamples of the conjecture can be ruled out.

Another example of the weak cosmic censorship was provided by Kegeles in 1978\cite{kegeles1978}. He showed that the gravitational collapse of a slowly rotating dust cloud forms a Kerr black hole, and this collapse does not radiate gravitational waves. This solution is consistent with the weak cosmic censorship. He also proposed an investigation of the rapidly rotating model to shed light on the weak cosmic censorship conjecture. 
%He added this investigation would be a generalization of the slowly rotating case.

In 1979, Kayll Lake introduced the relation between the third law of black hole thermodynamics and the weak cosmic censorship conjecture\cite{lake1979}. He stated that the impossibility of reducing the surface gravity to zero by finite steps implies that the GIC is not valid. He also emphasized that both the weak cosmic censorship conjecture and the third law of black hole thermodynamics invoke asymptotic flatness. He examined Reissner-Nordström-de Sitter spacetime with the third law and the weak cosmic censorship for a nonasymptotically flat spacetime. For this purpose, he studied the collapse model of a thin charged shell and investigated stabilities under perturbations to elucidate the physical realism of the model. He concluded that the meaning of the event horizon for this model and the thermodynamic meaning of the event horizon remarkably differ. If $\Lambda>0$, the third event horizon (this horizon is referred to as cosmological horizon in this paper) forms in addition to inner and outer event horizons of the Reissner-Nordström black hole. The possibility of the coalescence of the outer event horizon and third event horizon arises. If the third law of black hole thermodynamics is valid, we have degenerate configuration of event horizons, which is not expected. If a black hole has degenerate event horizons, this represents a zero temperature in terms of black hole mechanics.

In the same year, Farrugia and Hajicek proposed a counterexample of the third law of the black hole thermodynamics which does not contradict the assumption of the weak cosmic censorship conjecture\cite{farrugiahajicek1979}. They showed that the gravitational collapse of a spherically symmetric thin charged shell may lead to an extreme black hole. This process can be achieved in a finite amount of time; thus, it serves as a counterexample to the third law of black hole thermodynamics. Thus, their view is to not consider the third law of black hole thermodynamics as a general statement.

In 1979, Horowitz published a paper in which he introduced misconceptions about naked singularities and proposed two statements\cite{horowitz1979}. The first statement is as follows: \textquotedblleft The maximal development of every nonsingular, vacuum, asymptotically flat initial data set is an asymptotically flat spacetime". No counterexample exists for the first statement; however, this statement is strongly related to the concept of infinity. To address the drawback of the initial statement, he proposed another statement as follows: \textquotedblleft Let $\mathbf{M}$ be the maximal development of a noncompact vacuum initial data set $S$. If $\gamma\subset M$ is enlargeable, then $C1$ $[I^-(\gamma) \cap S]$ is noncompact."

\section{ The Developments of the WCCC in the 1980s}
\label{The Developments of the WCCC in the 1980s}
In 1980, Needham proposed a model\cite{needham1980} similar to Wald's approach. He investigated the interaction between a charged spinning test particle and an extreme Kerr-Newman black hole. Unlike Wald's 1974 work \cite{wald1974}, he proposed an arbitrary model of a spinning charged test particle, not assuming that the spinning test particle moves only in the equatorial plane or along the symmetry axis. His result was that the WCCC is valid if the test particle's parameters are small compared to those of the black hole. He also derived a condition. We refer to it as Needham's condition, which is given by \textcolor{white}{jjjjjkjbhgcvhgchgc}
\beq
\delta M- \Omega_H\delta J- \Phi_H\delta Q \geq 0,
\eeq
where $\Omega_H=\frac{a}{r_+^2+a^2}$ and $\Phi_H=\frac{Q\,r_+}{r_+^2+a^2}$. In this equation $r_+$ is the event horizon of a Kerr-Newman black hole given in Equation (\ref{eq:3.4}). This condition establishes a relation concerning the lower limit of energy required for a test particle or field to be absorbed by a black hole.

In 1981, Hiscock generalized Wald's 1974 work\cite{wald1974} to dyonic\footnote{having both eletric and magnetic charges} black holes and dyonic test particles\cite{hiscock1981}. He argued that the generic calculation showing that the black hole cannot be pushed beyond extremality remains valid if all parameters of the test particle, namely angular momentum, mass, electric and magnetic charges, are much smaller than the corresponding parameters of the black hole. He then identified four cases where this assumption is not valid, hence WCCC may be problematic:
\begin{enumerate}
\item a magnetically charged test particle into an extreme Reissner-Nordström black hole. 
\item a charged particle into an extreme Kerr black hole
\item a spinning test particle into an extreme Reissner-Nordström black hole
\item a test particle that possesses an orbital angular momentum into a Reissner-Nordström black hole
\end{enumerate}  
Then, he analyzed in some detail the cases of magnetic test particle and uniform spherical shell being captured by exremely electrically charged Reissner-Nordström black hole. The magnetically charged test particle $g$ with the mass $\mathbf{M}$ can be captured by an extreme Reissner-Nordström black hole if the following equation is satisfied:
\beq
m<\frac{g^2}{M}-\frac{3g^4}{2M^3}+\dots\quad. 
\eeq
For the shell case, he found that naked singularity cannot form as long as WEC is satisfied by the shell material. He then argued that the WEC should also be satisfied by the test particle.

%In 1981, Hiscock proposed a model\cite{hiscock1981} to extend Wald's approach with a magnetic charge. He studied the process of converting a black hole into a naked singularity by throwing a spinless test particle into the dyonic black hole. 
%He tried to achieve a final configuration with $M^2<a^2+q^2+p^2$, where $\mathbf{M}$ is the final mass of the black hole, $a$ is the final angular momentum per unit mass of the black hole, and $q$ and $p$ are the final electric and magnetic charges of the black hole, respectively. He demonstrated that a naked singularity cannot form; thus the WCCC cannot be violated.

%His model utilizes a Reissner-Nordström black hole, which is electrically charged, along with a test particle that is magnetically charged and possesses a high magnetic-to-mass ratio. This choice is considered the easiest case among them. He demonstrated  the existence of a naked singularity. The magnetically charged test particle $g$ with the mass $\mathbf{M}$ can be captured by an extreme Reissner-Nordström black hole if the following equation is satisfied:
%\begin{equation}
%m<\frac{g^2}{M}-\frac{3g^4}{2M^3}+\dots
%\end{equation}
%The approximate criteria were given above. To maintain the realistic solution, Hiscock also investigated a gravitational collapse scenario of spherical shell of magnetic monopoles into an electrically charged Reissner-Nordström black hole. By doing so, he completely solved the EFE to obtain the actual motion of the test particle.  His conclusion was that a naked singularity cannot form as long as the WEC is not violated by the test particle.

In the same year, Lake and Hellaby proposed a model for the occurrence of a naked singularity\cite{lakehellaby1981}. They studied the gravitational collapse of perfect fluid sphere in the Robertson-Walker spacetime. After the gravitational collapse of the sphere, the entire matter radiates and Vaidya spacetime forms. They demonstrated that a naked singularity occurs for this model. However, the conditions that satisfy the transition between spacetimes are not physically reasonable\footnote{Actually, Lake and Hellaby studied this model to form a naked singularity. There is no mechanism in GR for that transition.}, as there is a transition from isotropy to a flux of radiation radially. In addition to this, the collapse scenario includes a flux of created particles. Thus, this model is semi-classical model. In 1982, Lake generalized this model by changing initial conditions to be more realistic\cite{lake1982}. He did not use a perfect fluid sphere; instead, he used bulk viscosity as well. However, including bulk viscosity into the model did not alter the result of the previous work. He also commented that these models cannot be true counterexamples of the weak cosmic censorship conjecture since an accurate description of the energy-momentum tensor was not given.

In 1981, Glass and Harpaz studied the gravitational adiabatic collapse of a spherical body\cite{glassharpaz1981}. They demonstrate that a naked singularity does not occur if the DEC is satisfied.

In 1982, Nakamura and Sato published a gravitational collapse model for a non-rotating, axisymmetric star\cite{nakamurasato1982}. Their model used a mass equivalent to ten times that of the sun with a physically reasonable equation of state. If  the ratio of the initial internal energy of the star to its initial gravitational energy is less than $2/3$, this model demonstrates that a naked singularity forms\footnote{If the ratio greater than $2/3$, a Schwarzschild black hole forms.}. However, they emphasized that the condition which gives rise to a naked singularity might not be realistic, as the ratio can be close to one in a realistic collapse scenario.

In 1983, Newman proposed a new proof of the weak cosmic censorship conjecture mathematically\cite{newman1983}. He attempted to provide a mathematical definition of the concept of \textquotedblleft future asymptotic predictability". He emphasized that future asymptotic predictability is referred to as a form of the WCCC, since a naked singularity is the singularity that is visible from close to the future of a point. Here is the definition given by Newman: A weakly asymptotically simple and empty space-time ($\mathbf{M}$, $g_{\mu\nu}$) is future asymptotically predictable if the null convergence, the generic and strong causality conditions hold.

 On the other hand, Krolak proposed another proof of the weak cosmic censorship conjecture by introducing weakly asymptotically simple and empty spacetime\cite{krolak1983}. He used the notion of future asymptotic predictability and then tried to give a mathematical formulation of the weak cosmic censorship conjecture with strong causality condition as follows:

Let $\mathbf{M}$ and $g_{\mu\nu}$ be partially be partially future asymptotically predictable from a partial Cauchy surface $S$ and if the following conditions hold in ($\mathbf{M}$, $g_{\mu\nu}$), then ($\mathbf{M}$, $g_{\mu\nu}$) is future asymptotically predictable from $S$:
\begin{enumerate}
\item ($\mathbf{M}$, $g_{\mu\nu}$) is strongly causal
\item $R_{\mu\nu}k^{\mu}k^{\nu}\geq 0$ for every null vector $k^{\mu}$
\item Each NSTIP is a SSTIP\footnote{IP denotes indecomposable past set defined by Penrose and collaborators, T stands for terminal,  NSTIP and SSTIP are much more specialized sub-cases defined in \cite{krolak1983}, similary, $N\infty$-TIP is defined \cite{krolak1984}}
\item For any subset $V$ of $\mathbf{M}$ which is an IP such that $V\subset D^-(\mathscr{I}_0^+)$ and the intersection $D^+(S)\cap V$ is not empty, there exists a null geodesic generator $\lambda$ of $V$ in $D^+(S)$ such that $\lambda$ is an outgoing null geodesic.
\end{enumerate}

In 1984, Israel compared two conjectures: the weak cosmic censorship and Thorne's hoop conjecture\cite{israel1984}. The hoop conjecture\footnote{Israel referred to the hoop conjecture as the event horizon conjecture (EHC)} states that an event horizon forms under some circumstances: If we have a star of mass $M$ undergoing gravitational collapse, an event horizon forms if the circumference of the compacted region denoted by $C$ satisfies the condition $C\leq 2\pi(2GM/c^2)$. The EHC avoids referring to naked singularities; thus the difference between the EHC and the WCCC is significant. In this study, Israel believed that the EHC is likely correct, while the WCCC is not, as several counterexamples to the WCCC were found.

In 1984, Krolak gave a new mathematical definition to the weak cosmic censorship conjecture\cite{krolak1984}. He said that the WCCC holds if the spacetime is simple. Here is the definition of the simplicity of a spacetime:
 
Let $\mathbf{M}$ and $g_{\mu\nu}$ be a stably causal spacetime with $W$ being the topological sum of $N\infty$-TIPs\footnotemark[18] from a certain subfamily $\hat{W}$ of the family of all $N\infty$-TIPs in $\mathbf{M}$. Then, this spacetime is simple to the future of a partial Cauchy surface $S$ with respect to $\hat{W}$ if the set $I^+(S)\cap W$ is causally simple.

 In addition to this new definition, he said that this definition can be used for the cosmological models with a non-zero and negative cosmological constant $\Lambda$. In the same year, Kuroda demonstrated a counterexample to the weak cosmic censorship conjecture\cite{kuroda1984}. His model included a gravitational collapse of a spherical pure radiation in the Vaidya spacetime. He showed that shell focusing singularity exists for this model.

In 1984, Christodoulou demonstrated a counterexample of the weak cosmic censorship conjecture\cite{christodoulou1984}. He studied the behavior of light rays in the context of gravitational dust collapse\footnote{In fact, the OSD model analyzed the behavior of outgoing light rays, which introduced the idea of a black hole.}. This study focused on an inhomogeneous spherically symmetric model. In the future null infinity, light rays are redshifted without any restrictions on the initial density distribution.

In 1985, Tipler argued that the definitions and proofs proposed by Krolak \cite{krolak1983} and Newman \cite{newman1983} were invalid \cite{tipler1985}. He stated that Krolak's proof was incorrect due to mathematical errors and expressed doubt that these errors could be rectified. He criticized Krolak's choice of a partial Cauchy surface, noting that this choice artificially leads to breakdowns in global hyperbolicity. On the other hand, Newman failed because his assumption of spacetime curvature was not physically realistic; he employed a different notion for a strong curvature singularity. Tipler demonstrated that null geodesics with this condition do not fit within realistic spacetimes, such as open Friedmann spacetime and maximal Schwarzschild spacetime.

In 1987, Ori and Piran demonstrated that a naked singularity occurs in the gravitational collapse of a self-similar spherical adiabatic perfect fluid with the equation of state $p=c\,\rho$ if $c\ll1$\cite{oripiran1987}. They argued that this naked singularity resembles a shell-focusing singularity as a counterexample of the WCCC.

In 1987, Goldwirth and Piran investigated a numerical model\footnote{The theoretical model of this case was given by Christodoulou\cite{christodoulou1986}.} of the gravitational collapse of a spherical massless scalar field\cite{goldwirthpiran1987}. Altering initial conditions of the model, they calculated future evolutions of data. They concluded that a naked singularity cannot occur for the collapse scenario of the spherical massless scalar field. One year later, Abe studied the stability of this case theoretically\cite{abe1988}. For this model, we have a naked singularity if we include a scalar charge into the model. Abe demonstrated the instability in the scalar charge case. Thus, Abe's study validates the weak cosmic censorship conjecture.

In 1988, Roman proposed a model to test the validation of the third law of black hole thermodynamics\cite{roman1988}. He emphasized the connection between the weak cosmic censorship conjecture and the third law of black hole thermodynamics by arguing that violating the third law of black hole thermodynamics might imply destroying the event horizon of a black hole. He attempted to extract energy from the Reissner-Nordström black hole to reduce the surface gravity to zero in a finite time. He demonstrated that it is impossible to reduce surface gravity to zero in a finite time by extracting energy.

In the same year, Ori and Piran proposed a counterexample to the WCCC\cite{oripiran1988}. They used their 1987 model \cite{oripiran1987}, but this time they accepted $p\approx \rho$. They demonstrated that naked singularities exist for unbounded pressure values. They left the stability of the causal structure under perturbation as an open question.

At the end of 1988, Takashi Nakamura, Stuart L. Shapiro, and Saul A. Teukolsky published a paper\cite{nakamurashapiroteukolsky1988}. They solved the initial value problem in GR for axisymmetric prolate and oblate inhomogeneous dust spheroids. They increased the eccentricity of the spheroid to assess the hoop conjecture. With high eccentricity, naked singularities occur, and this is in agreement with the hoop conjecture. However, the validity of the WCCC is suspect.

In 1989, Roberts reviewed the solution of the EFE for the most general case of a static spherically symmetric massless scalar field\footnote{This solution is referred to as Wyman's solution. The causal structure of the Wyman spacetime is similar to the Reissner-Nordström spacetime with $q^2>M^2$.}. In that paper\cite{roberts1989}, he solved the EFE for a pure radiative massless scalar field (referred to as Vaidya Wyman spacetime). He showed that an event horizon does not exist for the Vaidya Wyman spacetime. Thus, he proposed a counterexample to the weak cosmic censorship conjecture.
\section{ The Developments of the WCCC in the 1990s}
\label{The Developments of the WCCC in the 1990s}
In 1990, Semiz\cite{semiz1990} unwittingly repeated the part about dyonic Kerr-Newman black holes of Hiscock's 1981 work. Using the electromagnetic duality transformation, he derived same result in the generic case, and eliminated some of the problematic cases identified in [19], but argued against that paper's conclusion in the main problematic case, that is, argued for preservation of  WCCC.

In the same year, Ori and Piran found the appearance of naked singularities in the spherical collapse scenario of a perfect fluid with a barotropic(isothermal) equation of state (i.e. $p=k\rho$ where $p$ is the pressure, $\rho$ is the total energy density and $k$ is a constant)\cite{oripiran1990}. For values of $k$ in the range $0 \leq k \leq 0.4$, there exist solutions that result in a naked singularity.

In 1991, Shapiro and Teukolsky provided a numerical computation for the relativistic model of a gravitational collapse\cite{shapiroteukolsky1991}. They investigated the collapse scenario for non-rotating, non-interacting homogeneous gas spheroids with both prolate and oblate shapes, considering various eccentricities and initial sizes. The matter is initially at rest; however, its velocity can later approach the speed of light. Their results suggest that a naked singularity can occur whereas the hoop conjecture is valid. 

In 1992, Joshi and Dwivedi analyzed the emergence of naked singularity solutions in self-similar gravitational collapse involving spherically symmetric perfect fluid with an adiabatic equation of state\cite{joshidwivedi1992}. They identified self-similar spacetimes that lead to the formation of naked singularities, demonstrating that these singularities arise when a mathematical term representing parameters, determined by the initial conditions, has positive real roots. They also emphasized that counter-examples to the WCCC given by this solution can be neglected if some energy conditions related to the positivity of energy are violated during the final stages of the collapse.

In 1997, Dwivedi and Joshi examined initial conditions for a gravitational collapse of dust cloud that forms a black hole or a naked singularity\cite{dwivedijoshi1997}. They emphasized that the choice of initial conditions determines the final fate of the gravitational collapse scenario. To investigate the final state of the collapse, they used the inhomogeneous dust collapse scenario with the free choice of initial velocities for a given matter distribution, satisfying the WEC. They demonstrated that the final state of a gravitational dust collapse can lead to the formation of a naked singularity or a black hole depending on initial data. Hence, they stated that there is no need for the WCCC.

\chapter{MODERN DEVELOPMENT OF THE WCCC IN CLASSICAL GENERAL RELATIVITY}
\label{Modern Development of the WCCC In Classical General Relativity}

In 1999, Veronika Hubeny proposed a thought experiment involving overcharging the Reissner-Nordström black hole by sending a test particle \cite{hubeny1999}. Unlike previous approaches, which following Wald\cite{wald1974} used extremal black holes, Hubeny started with a black hole close to extremal but not extremal. To neglect the back reaction effects, she sent a test particle whose mass, charge, and total energy are $M$, $q$, and $E$, respectively. Initially, the system satisfies the condition $m<E<q \ll Q<M$. The test particle falls radially toward the event horizon of the black hole, and the final configuration results in $Q + q > M + E$. In the final state, the total charge $Q_{\textit{final}}$ is $Q+q$ and the final mass $M_{\textit{final}}$ is less than $M+E$. Hence, for this problem, the Reissner Nordström black hole can be overcharged. The paper stated one helping factor is that some of the mass may be lost by gravitational radiation; and the backreaction effect is negligible. However, a numerical example provided in this paper demonstrates that the test particle might lose energy due to the backreaction effect and cannot be captured by the black hole. Thus, she emphasized that a detailed analysis must be done for even a single test particle to determine whether the WCCC is valid or not. Some later work\cite{bck2010} has argued the back reaction effect is not small enough to ignore the effect, in fact, helps to restore WCCC as also demonstrated by the numerical example referred to above. 

In 2010, Jacobson and Sotiriou extended the thought experiment involving the destruction of the event horizon of a Kerr-Newman black hole by incorporating a test body approximation in which the higher order contribution of the energy of the test body is not neglected\cite{jacobsonsotiriou2010}. They analyzed the possibility of overspinning a Kerr black hole, similar to Hubeny's work \cite{hubeny1999}. They demonstrated that naked singularities can form if a test particle is sent into a near-extreme Kerr black hole. However, this violation is not generic, it can be fixed by employing backreaction effect. Moreover, one year later, Saa and Santarelli demonstrated the possibility of overspinning a near extremal Kerr-Newman black hole\cite{saasantarelli2011}.

In 2011, Semiz proposed a gedanken experiment to investigate the WCCC\cite{semiz2011}. He attempted to push the extremality of a dyonic Kerr-Newman black hole by absorbing angular momentum and/or electric charge from a complex massive scalar field. He demonstrated that interaction between the field and black hole cannot lead to a naked singularity for any generic configuration of the field  and black hole parameters. This is the first work in the literature to perform a Wald type thought experiment with a test field instead of a test particle.

In 2013, Düztaş and Semiz investigated whether extremal or near extremal Kerr black hole can be overspun or not by bosonic fields\cite{duztassemiz2013}. They proposed a thought experiment by scattering massless scalar, electromagnetic, and gravitational fields from a Kerr black hole to investigate whether a naked singularity forms. They demonstrated that sending massless fields from a sub-extremal Kerr black hole can lead to a naked singularity if both self-force and gravitational radiational effects are neglected.

In 2015, Düztaş investigated a model aiming to challenge the WCCC\cite{duztas2015}. He proposed an interaction between a massless Dirac field and a Kerr black hole, demonstrating that such an interaction can lead to a naked singularity. This case serves as a counter-example to the WCCC. Additionally, he discussed the self-force and gravitational radiational effects, debating whether the naked singularity can be resolved. He concluded that the naked singularity cannot be prevented due to these effects.

The main reason for the different conclusions regarding the appearance of a leading naked singularity for fields with half-integer and integer spin is that a difference arises during the calculations of perturbation. For fields with integer spin, their energy-momentum tensor satisfies the WEC and NEC, while the energy-momentum tensor of fields with half-integer spin does not. 

The last decade has seen a large number of papers published on the WCCC, constructing Wald-type thought experiments aimed at validating or invalidating the conjecture, using a wide range of not really well motivated black hole solutions, or even black hole \emph{ans{\"a}tzse}; and fields or particles. The black holes considered include exotic ones such as BTZ, higher-dimensional, stringy, Taub-NUT, etc. and the fields/particles considered include again unverified ones such as dilatons, noncommutative fields.
\chapter{CONCLUSION}
\label{chapter:conclusion}

 Cosmic Censorship is probably the most important open problem in classical General Relativity\cite{penrose1999}. We discussed the beginnings of the idea around 1970, just after the singularity theorems showed the inevitability of development of singularities under some reasonable conditions in gravitational collapse. Singularities are obviously undesirable, and a threat to determinism in classical General Relativity, since that theory is time-reversal-symmetric: Anything that hits a singularity is expected to be destroyed, so what might come out of it? It seemed that nature should be protecting the universe from such dangerous beasts, hence the conjecture. There are mathematical solutions of Einstein's Equations featuring naked singularities which can affect the universe at large, so they cannot be forbidden outright. Instead, the conjecture assumes that \emph{if} the universe starts without naked singularities (the implicit assumption is that it must have), it cannot evolve them.
 
 In fact, this statement is the weak cosmic censorship conjecture, and is vague in a sense, since there is no generally accepted mathematical formulation for it. Naturally, this makes it hard to prove or disprove rigorously; instead, somewhat heuristic arguments are raised. One evidence of the validity of the weak cosmic censorship is the stability of the event horizon, which comprises the protection from the beast; e.g. Price showed that the linear perturbation for the spherically symmetric collapse to a Schwarzschild black hole is stable. Another angle of attack involves considering black hole solutions, whose mathematical representations —the metrics— for some ranges of parameters also represent naked singularities. Therefore it is natural to wonder if the parameters of these solutions could be pushed across that border by some process. That enterprise started with Wald's work in 1974, and continues to this day.
 
 Einstein Equations have the stress-energy-momentum tensor $T_{\mu\nu}$, i.e. matter on the right-hand-side, \emph{and General Relativity tells us nothing about the properties of matter}. Therefore it is conceivable that the (weak) cosmic censorship conjecture is really a statement —or conjecture— about matter, hence does not belong to General Relativity. In fact, several studies have shown/argued that WCCC is upheld if the matter satisfies some energy condition, but can be violated if that energy condition is violated.
 
 These could be construed as evidence against weak cosmic censorship, if one is inclined to not to put too much value in the energy conditions. In fact, there is more evidence against the weak cosmic censorship conjecture: Shell-focusing singularities seem to develop for both spherical and cylindrical collapse of perfect fluids, some numerical studies of initially prolate clouds suggest development of naked singularities, Wald-type thought experiments with incoming tailored fermion fields show WCCC violation, consistent with the energy-conditions discussion above.
 
Certain physicists hold the view that the validation of the weak cosmic censorship conjecture may ultimately come  through the advancements and insights of the exploration of quantum gravity. Edward Witten said that the weak cosmic censorship conjecture might be false in classical general relativity, but the version of the conjecture in string theory might be proven\cite{witten1991}. In any case, we have neither a good theory of Quantum Gravity yet, nor a convincing or testable String Theory, so the problem remains open. We have reviewed aspects of the problem, connections to related topics such as black hole thermodynamics, and properties of matter fields in this thesis. We expect similar exploration to continue in the foreseeable future.

\nocite{*}
 \bibliographystyle{styles/fbe_tez_v11.bst} %Still may have problems
\bibliography{references} %Still may have problems
%\flushleft
%\raggedright

%\appendix
%\chapter[AN APPENDIX TITLE THAT IS LONG AND THEREFORE\\
	%\hspace*{2.95cm} NEEDS MANUAL ADJUSTMENT IN LATEX CODE TO FIT\\
	%\hspace*{2.95cm} PROPERLY IN TABLE OF CONTENTS]{AN APPENDIX TITLE THAT IS LONG AND THEREFORE NEEDS MANUAL ADJUSTMENT IN LATEX CODE TO FIT PROPERLY IN TABLE OF CONTENTS}
	
%The appendices start here. After references section.

%\chapter{SAMPLE PAGES}
%This booklet (except its title page) is typeset in the
%format required for the theses. 

\end{document}